# Structural control of magnetic anisotropy in a strain driven multiferroic EuTiO$_3$ thin film

X. Ke[1,2], T. Birol[3], R. Misra[4], J.-H. Lee[5,6], B. J. Kirby[7], D. G. Schlom[6,8], C. J. Fennie[3], and J. W. Freeland[5]

[1]Department of Physics and Astronomy, Michigan State University, East Lansing, MI 48824, USA

[2] Neutron Sciences Directorate, Oak Ridge National Laboratory, Oak Ridge, TN 37831, USA

[3]School of Applied and Engineering Physics, Cornell University, Ithaca, NY 14853, USA

[4]Department of Physics and Materials Research Institute, Pennsylvania State University, University Park, Pennsylvania 16802, USA

[5]Advanced Photon Source, Argonne National Laboratory, Argonne, IL 60439, USA

[6]Department of Materials Science and Engineering, Cornell University, Ithaca, NY 14853, USA

[7]Center for Neutron Research, NIST, Gaithersburg, Maryland 20899, USA

[8]Kavli Institute at Cornell for Nanoscale Science, Ithaca, New York 14853, USA

Octahedral distortion plays a key role in engineering the physical properties of heterostructures composed of perovskite oxides. We observe a strong in-plane uniaxial magnetic anisotropy in a strain-enabled multiferroic EuTiO$_3$ thin film epitaxially grown on a (110)$_o$ DyScO$_3$ substrate. First principles calculations show that the magnetic anisotropy is closely correlated with the uniaxial TiO$_6$ octahedral tilting and the ferroelectric polarization of the film, indicating potential strong magnetoelectric coupling in the strain-engineered multiferroic system.

Epitaxial strain due to the lattice mismatch between an oxide film and the underlying substrate has been demonstrated to be an efficient strategy to tailor the physical properties of complex oxide heterostructures [1]. The prototypical strained-engineered materials are perovsktie oxides ($ABO_3$), where the strain accommodation through $BO_6$ octahedral rotations and distortions leads to changes of B-O-B bonding angles or B-O bonding length [2,3] and in the resultant materials properties. For instance, through the spin-lattice interaction, the epitaxial strain can have a profound impact on the magnetic anisotropy of magnetic perovskites, such as $La_{1-x}Sr_xMnO_3$ in which magnetic and electronic properties are determined by the Mn-O-Mn double exchange interaction and thus are sensitive to the epitaxial strain. Due to the magnetostriction effect, compressive strain generally results in an out-of-plane magnetic easy axis [4,5], while tensile strain results in an in-plane magnetic easy axis or a biaxial magnetic anisotropy ascribed to the dominant magnetocrystalline anisotropy [6,7,8]. In contrast, it was recently reported that biaxially compressively-strained $La_{0.67}Sr_{0.33}MnO_3$ films grown on $(110)_o$-oriented $NdGaO_3$ substrates [9] show strong in-plane uniaxial magnetic anisotropy, which are attributed to the anisotropic misfit stress relief through the lattice modulation along one of the in-plane orthogonal directions, i.e, $[100]_{pc}$ or $[010]_{pc}$, with the easy axis along the less compressively-strained direction. Note the subscripts "o" and "pc" represent orthorhombic and pseudocubic unit cell indices, respectively.

Recent efforts have been focused on the investigation, both qualitative and quantitative, of how the oxygen octahedral rotation induced by the epitaxial strain affect the functionalities of thin films [2,10,11,12]. For instance, scanning transmission electron microscopy has been exploited to characterize the interfacial octahedral rotation in the $BiFeO_3/La_{0.67}Sr_{0.33}MnO_3$ heterostructure [13] to elucidate the origin of the interfacial ferromagnetism [14]. For single epitaxial films, x-ray diffraction has used changes in the unit-cell size and local structure from x-ray absorption to infer changes in the local symmetry [15]. A direct observation of octahedral tilts is possible by examining half-order Bragg peaks as was recently shown for LaNiO$_3$ epitaxial films by comparing the intensity of half-order Bragg peaks with DFT calculated results [16]. Recent studies [15, 17] attempt to link together the octahedral rotations of the epitaxial films and the observed in-plane uniaxial magnetic anisotropy, which are both attributed to result from the different in-plane lattice constants of the orthorhombic substrate.

Most of the materials studied thus far have the magnetic ions occupying the B sites of the perovskite structure, which is naturally associated with the magnetic properties through BO$_6$ distortion. EuTiO$_3$, having Eu$^{2+}$ magnetic ions occupying the A site, has recently attracted intense attention as a multiferroic material [18,19]. EuTiO$_3$ is isostructural with the cubic SrTiO$_3$ at room temperature with the lattice constant of 3.905 Å, and in the bulk EuTiO$_3$ is a quantum paraelectric with G-type antiferromagnetic order below $T_N$ ~ 5.4 K [20,21]. Neutron, x-ray, and specific heat measurements reveal that EuTiO$_3$ undergoes a structural phase transition around 282 K [22,23,24] from cubic to tetragonal involving TiO$_6$ octahedral distortion [23,24,25], which is predicted to affect magnetic and electronic properties of EuTiO$_3$ [26, 27].

In this paper, we report the observation of in-plane uniaxial ferromagnetic anisotropy in EuTiO$_3$ commensurately strained in biaxial tension to (110)$_o$ DyScO$_3$ with the magnetic easy axis along one of the <110>$_{pc}$ pseudocubic EuTiO$_3$ axes. We show that this is closely correlated with the uniaxial TiO$_6$ octahedral tilting induced by the biaxial tensile strain and the resultant ferroelectric polarization, with both the rotation axis and polarization direction perpendicular to the magnetic easy axis. This suggests a strong magnetoelectric coupling in the strain-induced multiferroic EuTiO$_3$ films that could enable the control of ferromagnetism with an electric field.

EuTiO$_3$ films with a thickness of ~ 25 nm were grown using reactive molecular-beam epitaxy on (110)$_o$-oriented DyScO$_3$ substrates, which has an orthorhombic unit cell structure with lattice constants $a$ = 5.443 Å, $b$ = 5.717 Å, and $c$ = 7.901 Å [28]. DyScO$_3$ has a distorted perovskite structure and is often referred as pseudocubic with a lattice constant of about 3.947 Å. Thus, the EuTiO$_3$ film grown atop has a biaxial tensile strain of +1.1%. The orientation indices used hereafter are referenced with respect to the pseudocubic unit cell of the DyScO$_3$ substrate (see Fig. 2a). The excellent quality of the epitaxial EuTiO$_3$ film is evidenced by both x-ray diffraction high-resolution TEM measurements [19]. The commensurately-strained EuTiO$_3$ film was shown to be both ferromagnetic and ferroelectric, and detailed information about the sample growth has been previously described [19,29]. Magnetization properties of the EuTiO$_3$ film are studied using a Quantum Design superconducting quantum interference device (SQUID) and by polarized neutron reflectometry (PNR) measurements on NG1 reflectometer at the National Institute of Standards and Technology Center for Neutron Research.

PNR was used to probe the depth profiles of the sample's nuclear composition and in-plane vector magnetization [30]. Even though the Eu and Dy are strong neutron absorbers, grazing angle reflectivity is still experimentally feasible. An incident monochromatic neutron beam was polarized to be either parallel (spin up, or "+") or antiparallel (spin down, or "-") relative to the field. Using a spin analyzer positioned between the sample and detector, both non-spin-flip (++ and --) and spin-flip (+- and -+) reflectivities were measured. The difference of spin-up (++) and spin-down (--) reflectivities is dependent on the in-plane magnetization component parallel to the field direction, while the spin-flip signal is dependent on the component of the in-plane magnetization component perpendicular to the field. PNR measurements were conducted in a 10 Oe along the [010]$_{pc}$ direction (see inset of Fig. 1b) after cooling in 100 Oe along the same direction. The PNR data were corrected for background, beam footprint, and incident beam polarization.

The inset of Figure 1(a) shows the non-spin-flip PNR spectra taken at $T = 10$ K, which is above both the ferromagnetic transition of the film $T_c \sim 4.3$ K [19] and the antiferromagnetic transition temperature of DyScO$_3$ substrate $T_N \sim 3.1$ K [31]. In the paramagnetic regime above $T_c$, there is no difference between spin-up and spin-down reflectivity and model fitting of the spectra gives the thickness of the film to be 25 nm. The main panel of Fig. 1(a) shows the PNR spectra after the sample is field cooled into the ferromagnetic state ($T = 2.8$ K). The spin-up and spin-down reflectivities are well split, indicating a detectable projection of the magnetization along the field direction. The existence of spin-flip scattering signal represented by the blue triangles indicates a significant in-plane magnetic component perpendicular to the applied field direction, which leads us to conclude that the easy axis direction of the magnetic moment of the strained EuTiO$_3$ film at low temperature and zero field is not aligned along one of the edge directions, i.e., not along [100]$_{pc}$ or [010]$_{pc}$.

After trying simpler models, the PNR was fit using a model featuring a chemically uniform EuTiO$_3$ layer comprised of three distinct magnetic sublayers. This provided the best fit to the data, as illustrated by the solid curves in Fig. 1(a), which indicates a gradient in the magnetic state along the growth direction. The corresponding depth profile of both magnetization and its angle relative to the magnetic field direction are plotted in Fig. 1(b) and the schematic of the three-layer model is shown in the inset. The interface layer with a thickness of ~ 4 nm possesses the highest magnetization value (5.7 $\mu_B$ / Eu) followed by an intermediate layer with a magnetization of 4.5 $\mu_B$ / Eu and a thickness of about 16 nm. The top surface layer (~ 5 nm thick) has the smallest magnetization (2.0 $\mu_B$ / Eu) due to the non-magnetic Eu$^{3+}$ phase on the surface generated by exposure to air. Such non-uniformity in magnetization through the thickness is presumably attributable to the oxygen degradation of the uncapped sample in air as evidenced by TEM studies [32]. In addition, the obtained total magnetic moment of the sample, even for the interface region only, is smaller than the theoretical saturated magnetization (7 $\mu_B$/Eu) for $Eu^{2+}$, which is presumably also associated with the magnetic inhomogeneity as revealed by the recent magnetic force microscopy study [33]. The obtained magnetization direction is rotated by 45º relative to the $[010]_{pc}$ direction in the film plane as shown in the inset of Fig. 1(b), which directly indicates a uniaxial in-plane anisotropy. It is noteworthy that this magnetization direction is not sensitive to the models chosen for the data fitting.

To obtain a comprehensive picture of the magnetic anisotropy of the $EuTiO_3$ film, we also measured the temperature dependence of the remanent magnetization along different crystal orientations within ± 5º accuracy due to the square shape of the sample after it is cooled down from 10 K with a 100 Oe field (inset of Fig. 2b). Note that the magnetic signal of the $DyScO_3$ substrate (1 mm thick) at non-zero applied fields dominates over the $EuTiO_3$ film signal, making it difficult to extract hysteresis loops of the $EuTiO_3$ film. Two magnetic transitions are observed: one at $T_c$ ~ 4.3 K for the $EuTiO_3$ film and the other at $T_N$ ~ 3.1 K for the $DyScO_3$ substrate consistent with previous work on this system [19,31]. Owing to the in-plane tensile strain, the $EuTiO_3$ film has a smallest magnetization along the out-of-plane direction, $[001]_{pc}$, which is presumably associated with a magnetostriction effect due to the out-of-plane lattice contraction together with the shape anisotropy. The magnetization is the largest along the in-plane $[110]_{pc}$ direction in agreement with the neutron results. Along the orthorgonal direction, $[1\bar{1}0]_{pc}$, the magnetization is much smaller consistent with the uniaxial magnetic anisotropy seen in the PNR measurements. It is worth pointing out that, with the aid of x-ray diffraction, the in-plane $[110]_{pc}$ direction of the substrate corresponds to $[\bar{1}11]_o$ with orthorhombic indices.

Recent synchrotron x-ray diffraction measurements on biaxially strained $EuTiO_3/(110)_o$ $DyScO_3$ grown in the same batch as the sample in this study found by measuring various half integer Bragg reflections that there exists a single $TiO_6$ octahedral rotation domain in the $EuTiO_3$ film [34]. The film possesses a $a^-b^-c^0$ crystal symmetry pattern in Glazer notation [34,35], different from $a^0b^0c^-$ in a compressively strained $EuTiO_3/(100)$ LSAT film or from bulk (unstrained) $EuTiO_3$. In order to understand the correlation of such a peculiar octahedral rotation with the emergence of multiferroicity and magnetic anisotropy of the tensile strained $EuTiO_3$ film, we performed first principles density functional theory (DFT) calculations with spin-orbit coupling taken into account as implemented in Vienna *ab* initio simulation package (VASP) [36]. A plane wave cut-off of 500 eV and a 8 x 8 x 8 k-point grid for the 10-atom cell were used. Exchange-correlation energy was taken into account with the PBE functional [37].

While two symmetry patterns $a^-b^-c^0$ and $a^0b^0c^-$ compete in the bulk form of $EuTiO_3$, calculations [38] show that $a^-b^-c^0$ is energetically more favorable for $EuTiO_3$ under biaxial tension, as shown in Fig. 3(a). This is consistent with the crystal structure revealed by a recent synchrotron x-ray diffraction study [34]. In addition, due to the strong spin-lattice coupling, it is expected that the magnetic order of $EuTiO_3$ strongly depends on the structural state. DFT calculations indicate that the ferroelectric polarization is responsible for the observed ferromagnetism in $EuTiO_3$ films under tensile strain [18,19,39]. In Fig. 3(b) we plot the enthalpies of the ferromagnetic and antiferromagnetic states of $EuTiO_3$ under 1.1% tensile strain, which corresponds to $DyScO_3$, as a function of ferroelectric polarization. It is clearly seen that the antiferromagnetic state is favorable in the paraelectric structure, as well as in the structures with low polarization. For larger values of polarization, however, ferromagnetic order becomes energetically favorable, and for the fully relaxed ferroelectric structure the ferromagnetic state has an enthalpy that is ~0.25 meV/f.u. lower in energy than the antiferromagnetic state. Such a strong coupling between magnetism and polarization can be attributed to the recently proposed Ti-cation mediated exchange mechanism [40] which accounts for the antiferromagnetism observed in bulk EuTiO$_3$. The onset of the polarization displaces the Ti cations from the center of their oxygen coordination octahedra and decreases the associated antiferromagneitc exchange interaction. In other words, displacing the Ti cation has an extra energy cost in the antiferromagnetic state because the Ti-mediated exchange interaction favors the Ti to remain in the center of its oxygen coordination octahedra. Hence the ferromagnetic state becomes lower in energy for sufficiently large polarization.

Our DFT calculations also showed that the preferred direction of polarization is parallel to the octahedral rotation axis (in-plane diagonal directions, i.e, $[1\bar{1}0]_{pc}$) for EuTiO$_3$ under biaxial tension, which is consistent with a recent study [41]. In addition, to find out the correlation between the magnetic anisotropy and the TiO$_6$ octahedral rotation/polarization direction, we performed a noncollinear magnetism calculation. Interestingly, we found that the magnetization easy axis prefer energetically to align along $[111]_{pc}$ or $[11\bar{1}]_{pc}$ axes as shown in Fig. 4, compared to the other two cubic diagonal directions $[1\bar{1}1]_{pc}$ and $[1\bar{1}\bar{1}]_{pc}$ with an energy gain of 0.02 meV/f.u.. This implies that the ferroelectric polarization and magnetization vectors are perpendicular to each other, evidencing a strong magnetoelectric coupling in EuTiO$_3$ originating from the biquadratic $M^2P^2$ term [19]. Furthermore, we found that the preferred magnetization direction is different when we repeat the calculation using a structure without the ferroelectric polarization taken into account. This feature suggests that the magnetic anisotropy of multiferroic EuTiO$_3$ film is set *via* the coupling between the ferroelectric polarization and the ferromagnetic moment, both of which are correlated with the (nonmagnetic) TiO$_6$ octahedral rotation. This mechanism is distinct from others that account for the magnetic anisotropy reported in manganites by distorting the (magnetic) MnO$_6$ octahedra, i.e, changing the Mn-O-Mn bonding length/angle directly *via* epitaxial strain [9,15,17]. Instead, our observation in strained EuTiO$_3$ films is similar to recent reports in multiferroic BiFeO$_3$ films where FeO$_6$ octahedral rotations and polar distortions are strongly coupled. The weak canted ferromagnetization (~ 0.1 $\mu_B$/ Fe) in BiFeO$_3$ is induced by the Dzyaloshinskii- Moriya interaction perpendicular to the ferroelectric polarization [42,43]. Finally, comparison of the magnetic easy axis along the in-plane diagonal direction [110]$_{pc}$ observed in PNR and SQUID magnetometer measurements with an out-of-plane magnetization component calculated by DFT originates from the fact that the demagnetization energy term was not taken into account in DFT calculations, which due to the large Eu moment will lead to favoring an in-plane direction.

In summary, EuTiO$_3$ commensurately strained to (110)$_o$ DyScO$_3$ exhibits strong uniaxial magnetic anisotropy. The magnetic easy axis of this strain-enabled multiferroic is determined by magnetoelectric coupling between its ferroelectric polarization and its ferromagnetic moment both of which are correlated with the uniaxial TiO$_6$ octahedral distortion. Such a phenomenon is remarkably different from the magnetic anisotropy behavior regularly observed in epitaxial manganites. This study suggests the possibility of electric-field control of the magnetization of the strained enabled multiferroic EuTiO$_3$.

We are grateful for useful discussions with Prof. Peter E. Schiffer. X. K. acknowledges the support from the start-up funds at Michigan State University. Work at ORNL was supported by the Scientific User Facilities Division, Office of Basic Energy Sciences, DOE, and work at Argonne is supported by the U.S. Department of Energy, Office of Science, under Contract No. DE-AC02-06CH11357. T. B. and C. J. F. were supported by the DOE-BES under Grant No. DESCOO02334. R. M., J. H. L. and D.G.S were supported by the NSF MRSEC program (DMR-0820404).

FIGURE CAPTIONS

Figure 1. (a) Reflectivity as a function of $Q$ measured at $T$ = 2.8 K after the 25 nm thick EuTiO$_3$/(110)$_o$ DyScO$_3$ sample is cooled down in a 100 Oe magnetic field applied along the [001]$_o$ direction of the (110)$_o$ DyScO$_3$ substrate. The measurement field 10 Oe is applied to keep the neutron polarization direction. The inset shows the Q dependence of reflectivity of non spin-flip scattering measured at T = 10 K. Symbols are experimental data and solid curves are the fits, error bars correspond to ± 1 sigma; (b) Depth profile of the magnitude and angle of magnetization of the sample calculated from the data fitting. Inset shows the schematics of the film with nonuniform magnetization through the film thickness. Blue, violet, and red colors represent the top surface, the intermediate layer, and the interface respectively.

Figure 2: (a) Schematic of the crystal structure of EuTiO$_3$ under biaxial tensile strain with pseudocubic indices; (b) Temperature dependence of the magnetization of the EuTiO$_3$/DyScO$_3$ thin film with a 100 Oe magnetic field applied along various principle axes while cooling down from 10 K to 1.8 K. Measurements were performed at zero field during the warm up process. The inset shows the directions measured. Note that the pseudocubic directions of the epitaxial EuTiO$_3$ film are aligned with the pseudocubic directions of the DyScO$_3$ substrate.

Figure 3: (a) Calculated energies as a function of in-plane biaxial strain for polar structures with two different tilt symmetry patterns: $a^-b^-c^0$ and $a^0b^0c^-$; (b) Calculated enthalpies of antiferromagnetic (AFM) and ferromagnetic (FM) states as a function of the ferroelectric polarization for EuTiO$_3$ with 1.1% tensile strain, corresponding to commensurate EuTiO$_3$/(110)$_o$ DyScO$_3$.

Figure 4: Schematic illustrating that the calculated directions of the TiO$_6$ octahedral rotation axis (RA) and ferroelectric polarization (P) are perpendicular to the magnetic easy-axis (M) of the EuTiO$_3$ film. The directions given refer to the pseudocubic unit cell. Note that the demagnetization energy term was not taken into account in the DFT calculations, leading to the out-of-plane magnetization component.

Figure 1.

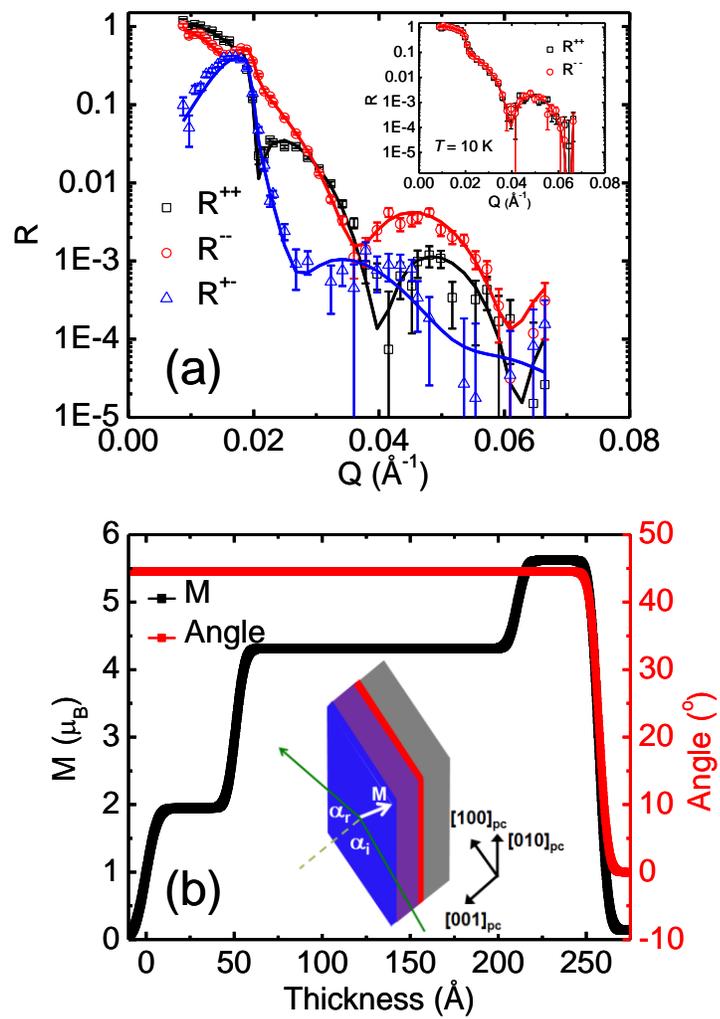

Figure 2.

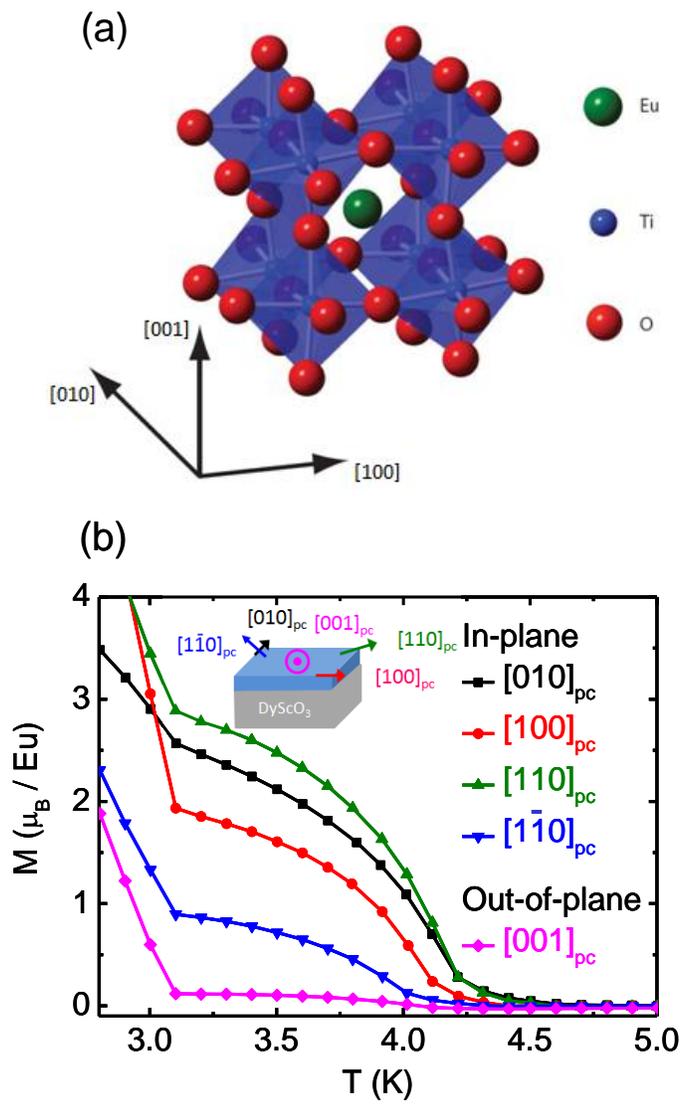

Figure 3.

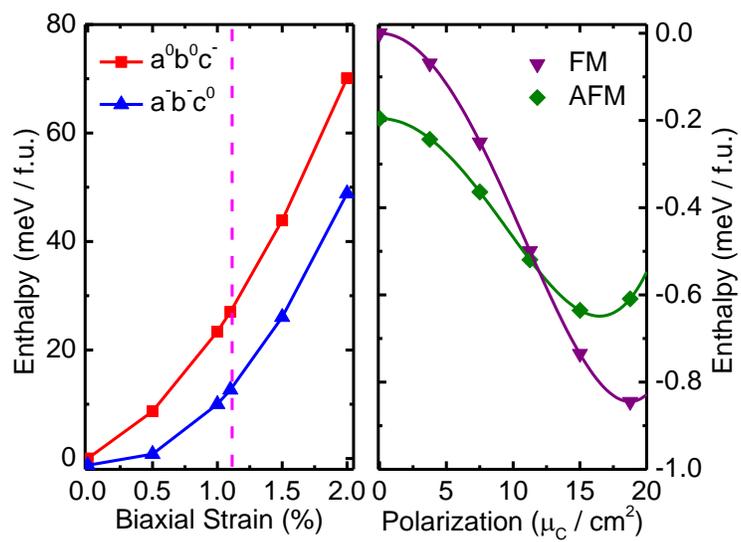

Figure 4.

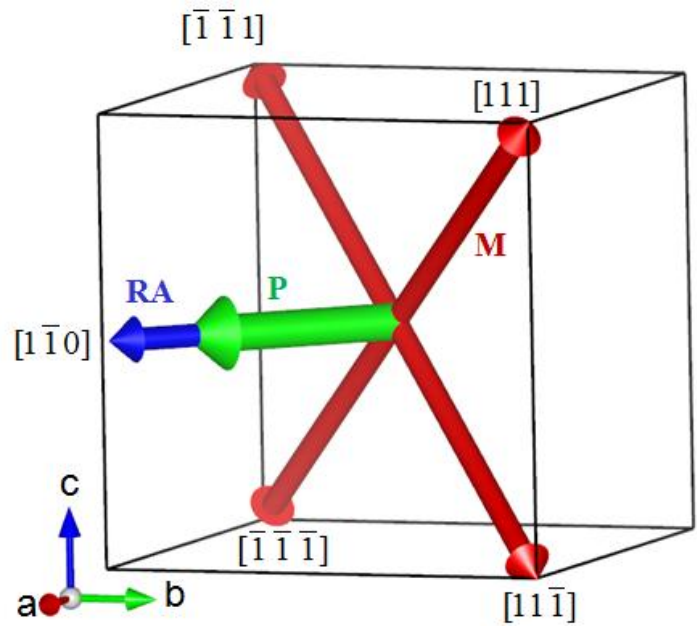